\documentclass[journal]{IEEEtran}
\usepackage{graphicx}
\usepackage{xspace}
\newcommand{\apenetp}{APEnet+\xspace}
\newcommand{\apelink}{APElink\xspace}

\newcommand{\nanet}{NaNet\xspace}
\newcommand{\nanetone}{\mbox{NaNet-1}\xspace}
\newcommand{\nanetten}{\mbox{NaNet-10}\xspace}
\newcommand{\nanetcube}{NaNet$^3$\xspace}
\newcommand{\gbe}{GbE\xspace}

\newcommand{\realtime}{\mbox{real-time}\xspace}
\newcommand{\lowlatency}{\mbox{low-latency}\xspace}
\newcommand{\rdma}{RDMA\xspace}
\newcommand{\nvidia}{NVIDIA\xspace}
\newcommand{\pcie}{PCIe\xspace}
\newcommand{\nios}{Nios~II\xspace}
\newcommand{\ie}{\textit{i.e.}\xspace}
\newcommand{\eg}{\textit{e.g.}\xspace}
\newcommand{\tengbe}{10-GbE\xspace}
\newcommand{\cerenkov}{\v{C}erenkov\xspace}
\newcommand{\km}{KM3NeT-IT\xspace}
\newcommand{\de}{DE5-net\xspace}

\newcommand{\mbps}{Mbps\xspace}
\newcommand{\gpudirect} {GPUDirect\xspace}
\newcommand{\altera} {Altera\xspace}

\DeclareGraphicsExtensions{.pdf,.jpeg,.png}

\begin{document}
\title{\nanet: a \mbox{Low-Latency}, \mbox{Real-Time}, \mbox{Multi-Standard}
  Network Interface Card with GPUDirect Features}
%
%

\author{A.~Lonardo\IEEEauthorrefmark{1}, 
  F.~Ameli\IEEEauthorrefmark{1}, 
  R.~Ammendola\IEEEauthorrefmark{2}, 
  A.~Biagioni\IEEEauthorrefmark{1}, 
  O.~Frezza\IEEEauthorrefmark{1},
  G.~Lamanna\IEEEauthorrefmark{3}\IEEEauthorrefmark{4}, 
  F.~Lo~Cicero\IEEEauthorrefmark{1},
  M.~Martinelli\IEEEauthorrefmark{1},
  C.~Nicolau\IEEEauthorrefmark{1}, 
  P.~S.~Paolucci\IEEEauthorrefmark{1}, 
  E.~Pastorelli\IEEEauthorrefmark{1}, 
  L.~Pontisso\IEEEauthorrefmark{5}, 
  D.~Rossetti\IEEEauthorrefmark{6},
  F.~Simeone\IEEEauthorrefmark{1}, 
  F.~Simula\IEEEauthorrefmark{1}, 
  M.~Sozzi\IEEEauthorrefmark{3}\IEEEauthorrefmark{4} ,
  L.~Tosoratto\IEEEauthorrefmark{1}, 
  and P.~Vicini\IEEEauthorrefmark{1}\\


  \thanks{\IEEEauthorrefmark{1}INFN Sezione di Roma,~Italy.}%
  \thanks{\IEEEauthorrefmark{2}INFN Sezione di Tor Vergata,~Italy.}%
  \thanks{\IEEEauthorrefmark{3}INFN Sezione di Pisa,~Italy.}%
  \thanks{\IEEEauthorrefmark{4}CERN,~Switzerland.}%
  \thanks{\IEEEauthorrefmark{5}Universit\`a di Roma Sapienza, Dipartimento di Fisica,~Italy.}%
  \thanks{\IEEEauthorrefmark{6}NVIDIA Corporation,~U.S.A.}
  \thanks{A.~Lonardo is the corresponding author (\mbox{alessandro.lonardo@roma1.infn.it}.)}%
  \thanks{This work was supported in part by the EU Framework Programme 7
    EURETILE project, grant number 247846; R. Ammendola was supported by
    MIUR (Italy) through the INFN SUMA project; G.~Lamanna and M.~Sozzi thank
    the GAP project, partially supported by MIUR under grant RBFR12JF2Z
    ``Futuro in ricerca 2012''.}
  \thanks{Manuscript received May 22, 2014.}%
}

\maketitle
\thispagestyle{empty}

\begin{abstract}
While the GPGPU paradigm is widely recognized as an effective approach
to high performance computing, its adoption in \mbox{low-latency},
\realtime systems is still in its early stages.

Although GPUs typically show deterministic behaviour in terms of
latency in executing computational kernels as soon as data is
available in their internal memories, assessment of \realtime features
of a standard GPGPU system needs careful characterization of all
subsystems along data stream path.
The networking subsystem results in being the most critical one in
terms of absolute value and fluctuations of its response latency.

Our envisioned solution to this issue is \nanet, an \mbox{FPGA-based}
\pcie Network Interface Card (NIC) design featuring a configurable and
extensible set of network channels with direct access through
\gpudirect to NVIDIA Fermi/Kepler GPU memories.

\nanet design currently supports both standard - \gbe (1000BASE-T) and
\tengbe (10Base-R) - and custom - 34~Gbps \apelink and 2.5~Gbps
deterministic latency KM3link - channels, but its modularity allows
for a straightforward inclusion of other link technologies.

To avoid host OS intervention on data stream and remove a possible
source of jitter, the design includes a network/transport layer
offload module with \mbox{cycle-accurate}, \mbox{upper-bound} latency,
supporting UDP, KM3link Time Division Multiplexing and \apelink
protocols.

After \nanet architecture description and its latency/bandwidth
characterization for all supported links, two real world use cases
will be presented: the \mbox{GPU-based} low level trigger for the RICH
detector in NA62 experiment at CERN and the \mbox{on-/off-shore} data
link for KM3 underwater neutrino telescope.



Results of \nanet performances in both experiments will be reported
and discussed.
\end{abstract}



\section{\nanet design overview}
\label{sec:nanetone}
%
\IEEEPARstart{N}{aNet} is a modular design of a \lowlatency \pcie
\rdma NIC supporting different network links, namely standard \gbe
(1000BASE-T) and \tengbe (10Base-R), besides custom 34~Gbps
\apelink~\cite{APEnetTwepp:2013} and 2.5~Gbps deterministic latency
optical KM3link~\cite{Aloisio:2011:NSS}.
The design includes a network stack protocol offload engine yielding a
very stable communication latency, a feature making \nanet suitable
for use in \realtime contexts; \nanet \gpudirect \rdma capability,
inherited from the \apenetp 3D torus NIC dedicated to HPC
systems~\cite{APEnetChep:2012}, extends its \mbox{realtime-ness} into
the world of GPGPU heterogeneous computing.

\begin{figure}[!htb]
  \centering
  \includegraphics[trim=60mm 35mm 60mm 20mm, clip,width=.49\textwidth]{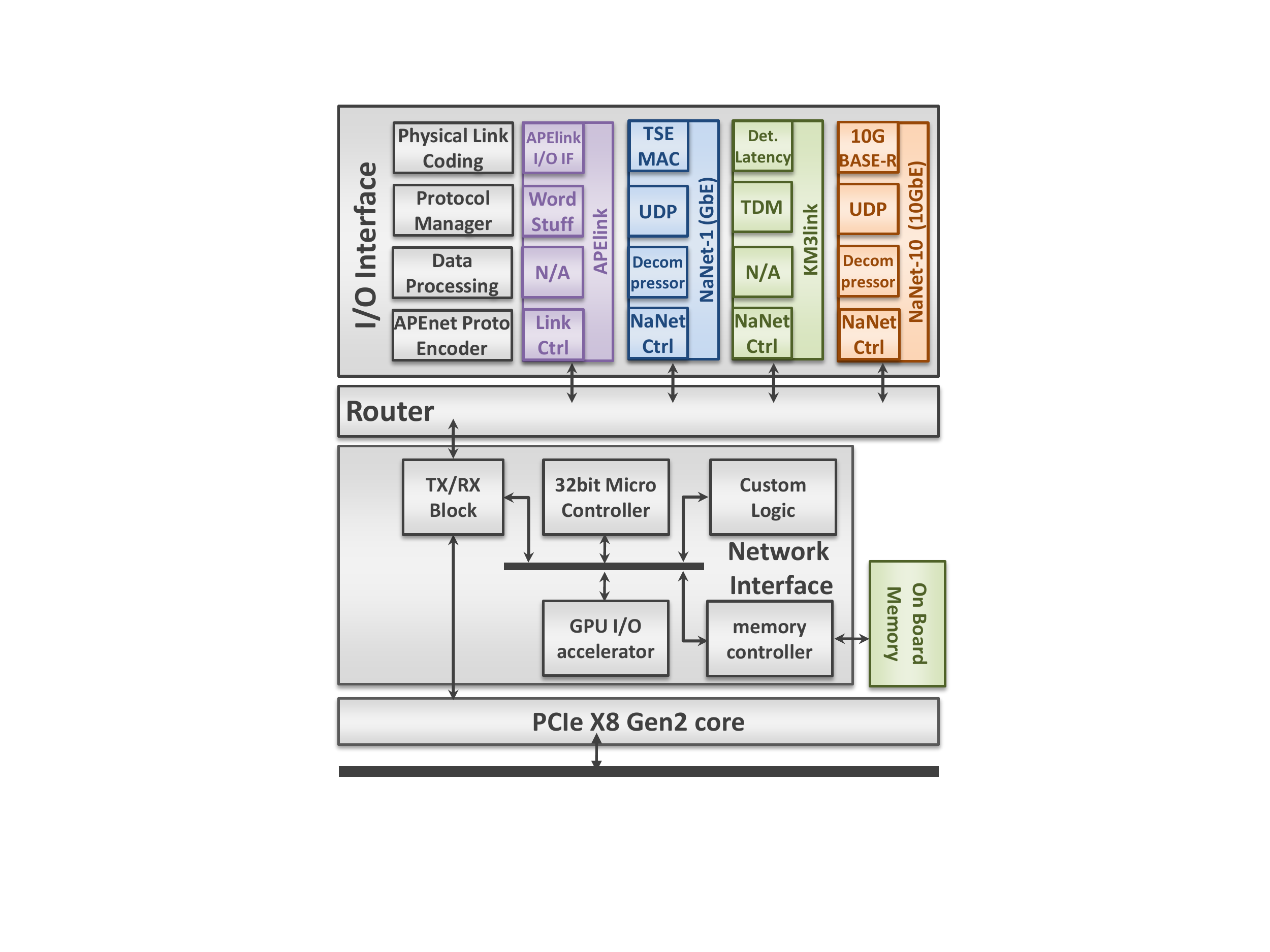}
  \caption{\nanet architecture schematic.}
  \label{fig:NaNet}
\end{figure}

\nanet design is partitioned into 4 main modules: \textit{I/O
  Interface}, \textit{Router}, \textit{Network Interface} and
\textit{\pcie Core} (see Fig.~\ref{fig:NaNet}).

I/O Interface module performs a \mbox{4-stages} processing on the data
stream: following the OSI Model, the Physical Link Coding stage
implements, as the name suggests, the channel physical layer (\eg
1000BASE-T) while the Protocol Manager stage handles, depending on the
kind of channel, data/network/transport layers (\eg Time Division
Multiplexing or UDP); the Data Processing stage implements application
dependent transformations on data streams (\eg performing
compression/decompression) while the APEnet Protocol Encoder performs
protocol adaptation, encapsulating inbound payload data in \apelink
packet protocol, used in the inner \nanet logic, and decapsulating
outbound \apelink packets before \mbox{re-encapsulating} their payload
in output channel transport protocol (\eg UDP).

The Router module supports a configurable number of ports implementing
a full crossbar switch responsible for data routing and dispatch.
Number and \mbox{bit-width} of the switch ports and the routing
algorithm can each be defined by the user to automatically achieve a
desired configuration.
The Router block dynamically interconnects the ports and comprises
a fully connected switch, plus routing and arbitration blocks 
managing multiple data flows @2.8~GB/s

The \textit{Network Interface} block acts on the trasmitting side by gathering
data coming in from the \pcie port and forwarding them to the
Router destination ports while on the receiving side it provides support for
\rdma in communications involving both the host and the GPU (via the
dedicated \textit{GPU I/O Accelerator} module). A \nios
$\mu$controller in included to support configuration and runtime operations.


Finally, the \pcie Core module is built upon a powerful commercial core from PLDA
that sports a simplified but efficient backend interface and multiple DMA engines.

As will be shown in the following, this general architecture has been
specialized to be employed in several contexts, and implemented on
several devices: \altera Stratix IV and V FPGA development kit and
Terasic \de board.

\section{\nanetone: a NIC for the NA62 \mbox{GPU-based} low level trigger}
  
The NA62 experiment at CERN~\cite{Lamanna:2011zz} aims at measuring
the Branching Ratio of the \mbox{ultra-rare} decay of the charged 
Kaon into a pion and a \mbox{neutrino-antineutrino} pair.
The NA62 goal is to collect $\sim100$ events with a signal to 
background ratio 10:1, using a novel technique with a \mbox{high-energy} 
(75~GeV) unseparated hadron beam decaying in flight.
In order to manage the \mbox{high-rate} data stream due to a
$\sim$ 10~MHz rate of particle decays illuminating the detectors,
a set of trigger levels will have to reduce this rate by three
orders of magnitude.
The entire trigger chain works on the main digitized data
stream~\cite{Avanzini:2010zz}.

The low-level trigger (L0), implemented in hardware by means of FPGAs
on the readout boards, reduces the data stream by a factor 10 to meet
the maximum design rate for event readout of 1~MHz.
The upper trigger levels (L1 and L2) are 
\mbox{software-implemented} on a commodity PC farm for further
reconstruction and event building. 

In the standard implementation, the FPGAs on the readout boards
compute simple trigger primitives on the fly, such as hit
multiplicities and rough hit patterns, which are then
\mbox{time-stamped} and sent to a central processor for matching and
trigger decision.
Thus the maximum latency allowed for the synchronous L0 trigger is related
to the maximum data storage time available on the data acquisition
boards.
For NA62 this value is up to 1~ms, in principle allowing use of more
compute demanding implementations at this level, \ie the GPUs. 

As a first example of GPU application in the NA62 trigger system we
studied the possibility to reconstruct rings in the RICH.
The RICH L0 trigger processor is a \lowlatency
synchronous level and the possibility to use the GPU must be verified.
In order to test feasibility and performances, as a starting point 5
algorithms for single ring finding in a sparse matrix of 1000 points 
(centered on the PMs in the RICH spot) with 20 firing PMs (``hits'') 
on average have been implemented.
Results of this study are available in~\cite{Collazuol:2012zz} and
show that GPU processing latency is stable and reproducible once data
are available in the device internal memory.

In order to fully characterize latency and throughput of the
\mbox{GPU-based} RICH L0 trigger processor (GRL0TP), data communication
between the detector readout boards (TEL62) and the L0 trigger
processor (L0TP) need to be kept under control.
The requisite on bandwidth is 400$\div$700~MB/s, depending on the
final choice of the primitives data protocol which in turn depends on
the amount of preprocessing actually to be implemented in the TEL62
FPGA.
Therefore, in the final system 4$\div$6 \gbe links will be used to
extract primitives data from the readout board towards the L0TP.

The \nanetone NIC was integrated in the GRL0TP prototype, using the
``system loopback'' setup described in section~\ref{sec:perf}.

\subsection{\nanetone implementation}
\label{sec:hw}

\begin{figure}[!htb]
  \centering
  \includegraphics[trim=10mm 10mm 10mm 10mm, clip,width=.49\textwidth]{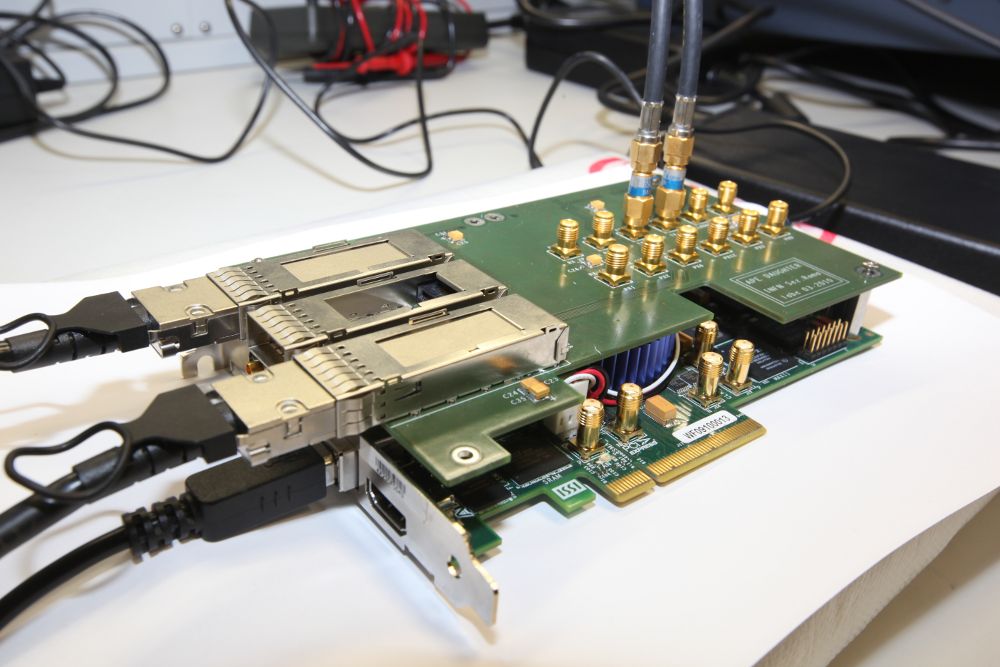}
  \caption{\nanetone on \altera Stratix IV dev. board EP4SGX230KF40C2 with custom
    mezzanine card + 3 \apelink channels.}
  \label{fig:board}
\end{figure}

The \nanetone is a \pcie Gen2 x8 NIC featuring a standard \gbe
interface implemented on \altera Stratix IV FPGA
Development Kit (see Fig.~\ref{fig:board}).  
A custom mezzanine mounting 3 QSFP+ connectors, was designed to be
optionally mounted on top of the \altera board and makes \nanetone
able to manage 3 \mbox{bi-directional} \apelink channels with 
switching capabilities up to 34~Gbps.
\apelink adopts a proprietary data transmission word stuffing
protocol; this is pulled for free into \nanetone.

The \gbe transmission system follows the general I/O interface 
architecture description of Fig.~\ref{fig:NaNet}.

The \altera Triple Speed Ethernet Megacore (TSE MAC) is the 
Physical Link Coding, providing complete 10/100/1000~Mbps Ethernet IP
modules.
%
%

The \textit{UDP Offloader} analyzes and interprets the data protocol.
It deals with UDP packets payload extraction and provides a
\mbox{32-bit} wide channel achieving 6.4~Gbps which is a 6 times
greater bandwidth than what the standard \gbe requires.

The data coming from TSE MAC are collected by the UDP Offloader
through the \altera Avalon Streaming Interface and redirected into the
\nanet hardware processing data path, avoiding the use of the the FPGA
\mbox{on-board} $\mu$controller (\nios) from UDP traffic management.
%
%

The \textit{\nanet Controller} translates the \mbox{UDP-encapsulated}
data packets into \apenetp encapsulated ones, then hands them over to
the Network Interface that takes care of moving them to their GPU
memory buffer destination.
%
%

%

\subsection{Results}
\label{sec:perf}

\nanetone performances were assessed on a Supermicro SuperServer
\mbox{6016GT-TF}.
The setup comprised a \mbox{X8DTG-DF} (Tylersburg chipset --- Intel
5520) dual socket motherboard, 2 Intel 82576 \gbe ports and \nvidia
M2070 GPU; sockets were populated with Intel Xeon X5570 @2.93~GHz.

The host simulates the RO board by sending UDP packets containing
primitives data from the host system \gbe port to the \gbe port hosted
by \nanetone, which in turn streams data directly towards CLOPS
in GPU memory that are sequentially consumed by the CUDA kernel
implementing the ring reconstruction algorithm.
This measurement setup is called ``system loopback''.

Exploiting the x86 Time Stamp Counter (TSC) register as a common
time reference, it was possible in a single process test application
to measure latency as time difference between when a received buffer
is signalled to the application and the moment before the first UDP
packet of a bunch (needed to fill the receive buffer) is sent through
the host \gbe port.
Communication and kernel processing tasks were serialized in order to
perform the measure; 
This represents a \mbox{worst-case} situation: given \nanetone RDMA
capabilities, during normal operation this serialization does not
occur and kernel processing seamlessly overlaps with data transfer.
Similarly, we closed in a loopback configuration two of the three 
available \apelink ports and performed the same measurement.
\begin{figure}[!htb]
  \includegraphics[trim=20mm 20mm 10mm 15mm,clip,width=.49\textwidth]{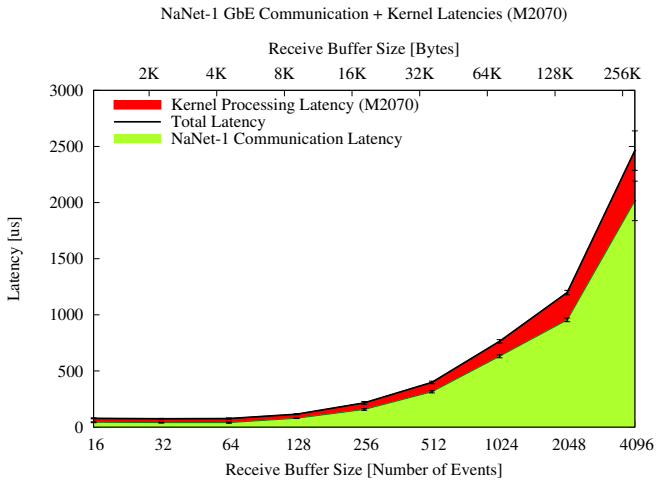}
  \caption {Latency of \nanetone \gbe data transfer and of ring
    reconstruction CUDA kernel processing.}
  \label{fig:latenza_nanet_fermi}
\end{figure}
\begin{figure}[!htb]
  \includegraphics[trim=20mm 20mm 10mm 15mm,clip,width=.49\textwidth]{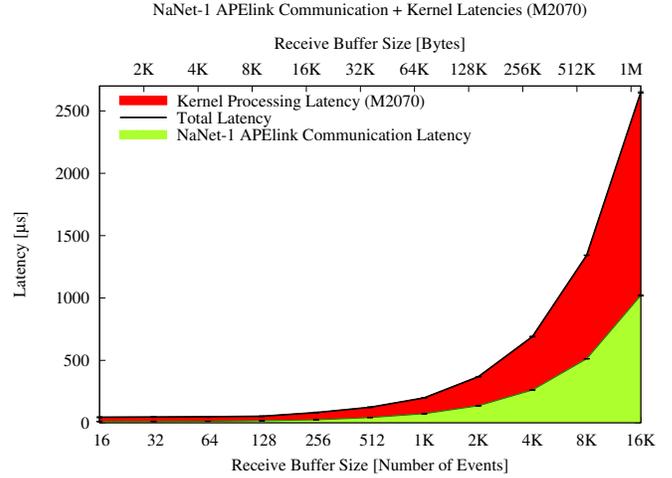}
  \caption {Latency of \nanetone \apelink data transfer and of ring
    reconstruction CUDA kernel processing.}
  \label{fig:latenza_apelink}
\end{figure}
In Fig.~\ref{fig:latenza_nanet_fermi} latencies for varying size
buffer transfers in GPU memory using the \gbe link are represented.
Besides the smooth behaviour increasing receive buffer sizes,
fluctuations are minimal, matching both constraints for \realtime and,
compatibly with link bandwidth, \lowlatency on data transfers; for a
more detailed performance analysis,
see~\cite{NanetTwepp:2013}.

Bandwidth and latency performances for \nanetone \apelink channel are
in Fig.~\ref{fig:latenza_apelink}.
It is clear that the system remains within the 1~ms time budget 
with GPU receive buffer sizes in the $128\div1024$~events range.
Although real system physical link and data protocol were used to show
the \realtime behaviour on \nanetone, we measured on a reduced
bandwidth single \gbe port system that could not match the
10~MEvents/s experiment requirement for the GRL0TP.

To demonstrate the suitability of \nanetone design for the
\mbox{full-fledged} RICH L0TP, we decided to perform equivalent
benchmarks using one of its \apelink ports instead of the \gbe one.
Current implementation of \apelink is able to sustain a data flow up
to $\sim20$~Gbps.
Results for latency of the \mbox{\apelink-fed} RICH L0TP are shown in 
Fig.~\ref{fig:latenza_apelink}: a single \nanetone \apelink data
channel between RICH RO and GRL0TP systems roughly matches trigger
throughput and latency requirements for receiving buffer size in the
$4\div5$~Kevents range.

\section{\nanetcube: the \mbox{on-shore} readout and \mbox{slow-control} board for the \km underwater neutrino telescope}


\km is an underwater experimental apparatus for the detection of high
energy neutrinos in the TeV$\div$PeV range based on the \cerenkov
technique.

The detector measures the visible \cerenkov photons induced by
charged particles propagating in sea water at speed larger than that
of light in the medium, and consists of an array of photomultipliers
(PMT).

The charged particle track can be reconstructed measuring the time of
arrival of the \cerenkov photons on the PMTs, whose positions must be
known.

The \km detection unit is called \textit{tower} and consists of
14~floors vertically spaced 20~meters apart.
The floor arms are about 8~m long and support 6 glass spheres called
Optical Modules (OM): 2 OMs are located at each floor end and 2 OMs in
the middle of the floor; each OM contains one 10~inches~PMT and the
\mbox{front-end} electronics needed to digitize the PMT signal, format
and transmit the data.
Each floor hosts also two hydrophones, used to reconstruct in
\realtime the OM position, and, where needed, oceanographic
instrumentation to monitor site conditions relevant for the detector.

All data produced by OMs, hydrophones, and instruments, are collected
by an electronic board contained in a vessel at the centre of the
floor; this board, called \textit{Floor Control Module} (FCM) manages
the communication between the \mbox{on-shore} laboratory and the
underwater devices, also distributing the timing information and
signals.
Timing resolution is fundamental in track reconstruction, \ie pointing
accuracy in reconstructing the source position in the sky.
An overall time resolution of about 3~ns yields an angular resolution
of 0.1~degrees for neutrino energies greater than 1~TeV.
Such resolution depends on electronics but also on position
measurement of the OMs, which is, in fact, continuously tracked.

The spatial accuracy required should be better than 40~cm.

\subsection{The \km DAQ and data transport architecture}
\label{sec:DAQArc}
The DAQ architecture is heavily influenced by the need of a common
timing distributed all over the system in order to correlate signals
from different parts of the apparatus with the required nanosecond
resolution.
The aim of the data acquisition and transport electronics is to label
each signal with a ``time stamp'', \ie the hit arrival time, in order
to reconstruct tracks.
This need implies that the readout electronics, which is spatially
distributed, require common timing and a known delay with respect to a
fixed reference.
The described constraints hinted to the choice of a synchronous link
protocol which embeds clock and data with a deterministic latency; due
to the distance between the apparatus and shoreland, the transmission
medium is forced to be an optical fiber.

All floor data produced by the OMs, the hydrophones and other devices
used to monitor the apparatus status and the environmental conditions,
are collected by the Floor Control Module (FCM) board, packed together
and transmitted through the optical link.
Each floor is independent from the others and is connected by an
optical bidirectional virtual \mbox{point-to-point} connection to the
\mbox{on-shore} laboratory.

The data stream that a single floor delivers to shore has a rate of
$\sim$300~\mbps, while the \mbox{shore-to-underwater} communication
data rate is much lower, consisting only of \mbox{slow-control} data
for the apparatus.
To preserve optical power budget, the link speed is operated at
800~\mbps, which, using an 8B10B encoding, accounts for a 640~\mbps of
user payload, well beyond experimental requirement.

Each FCM needs an \mbox{on-shore} communication endpoint counterpart.
The limited data rate per FCM compared with \mbox{state-of-the-art}
link technologies led us to designing \nanetcube, an \mbox{on-shore}
readout board able to manage multiple FCM data channels.

This design represents a \nanet customization for the \km experiment,
adding support in its I/O interface for a synchronous link protocol
with deterministic latency at physical level and for a Time Division
Multiplexing protocol at data level (see Fig.~\ref{fig:NaNet}).

\subsection{\nanetcube implementation}
The first stage design for \nanetcube was implemented on an evaluation
board from Terasic, the \de board, which is based on \altera
Stratix-V~GX FPGA, supports up to 4~SFP+ channels and a \pcie x8 edge
connector.

The first constraint to be satisfied requires having a time delta with
nanosecond precision between the wavefronts of three clocks:
\begin{itemize}
\item the first clock is an \mbox{on-shore} reference one (typically
  coming from a GPS and redistributed by custom fanout boards) and is
  used for the optical link transmission from \nanetcube towards the
  underwater FCM;
\item the second clock is recovered from the incoming data stream by a
  CDR module at the receiving end of the FCM which uses it for sending
  its data payload from the apparatus back \mbox{on-shore};
\item a third clock is again recovered by \nanetcube while decoding
  this payload at the end of the loop.
\end{itemize}
The link established in this way is fully synchronous.

The second fundamental constraint is the deterministic latency that
the \altera Stratix device must enforce --- as the FCM does --- on both
forward and backward path to allow correct time stamping of events on
the PMT.

In this way, the \nanetcube board plays the role of a bridge between
the 4 FCMs and the FCMServer --- \ie the hosting PC --- through the
\pcie bus.
Control data en route to the underwater apparatus are correctly sent
over the \pcie bus to the \nanetcube board, which then routes the data
to the required optical link.
On the opposite direction, both control and hydrophones data plus
signals from the \mbox{front-end} boards are extracted from the
optical link and \mbox{re-routed} on the \pcie bus towards an
application managing all the data.
Since the data rate supported by the \pcie bus is much higher than the
data produced by the \mbox{off-shore} electronics, we foresee to
develop a custom board supporting more than 4 optical links.
On the other hand, the \gpudirect \rdma features of \nanet, fully
imported in \nanetcube design, will allow us, at a later stage, to
build an effective, \realtime, \mbox{GPU-based} platform, in order to
investigate improved trigger and data reconstruction algorithms.

At a higher level, two systems handle the data that come from and go
to \mbox{off-shore}: the Trigger System, which is in charge of
analysing the data from PMTs extracting meaningful data from noise,
and the \mbox{so-called} Data Manager, which controls the apparatus.
The FCMServer communicates with these two systems using standard
\tengbe network links.

\subsection{The \nanetcube preliminary results}
\begin{figure}[!htb]
  \centering
  \includegraphics[trim=60mm 35mm 60mm 20mm, clip, angle=270, width=.49\textwidth]{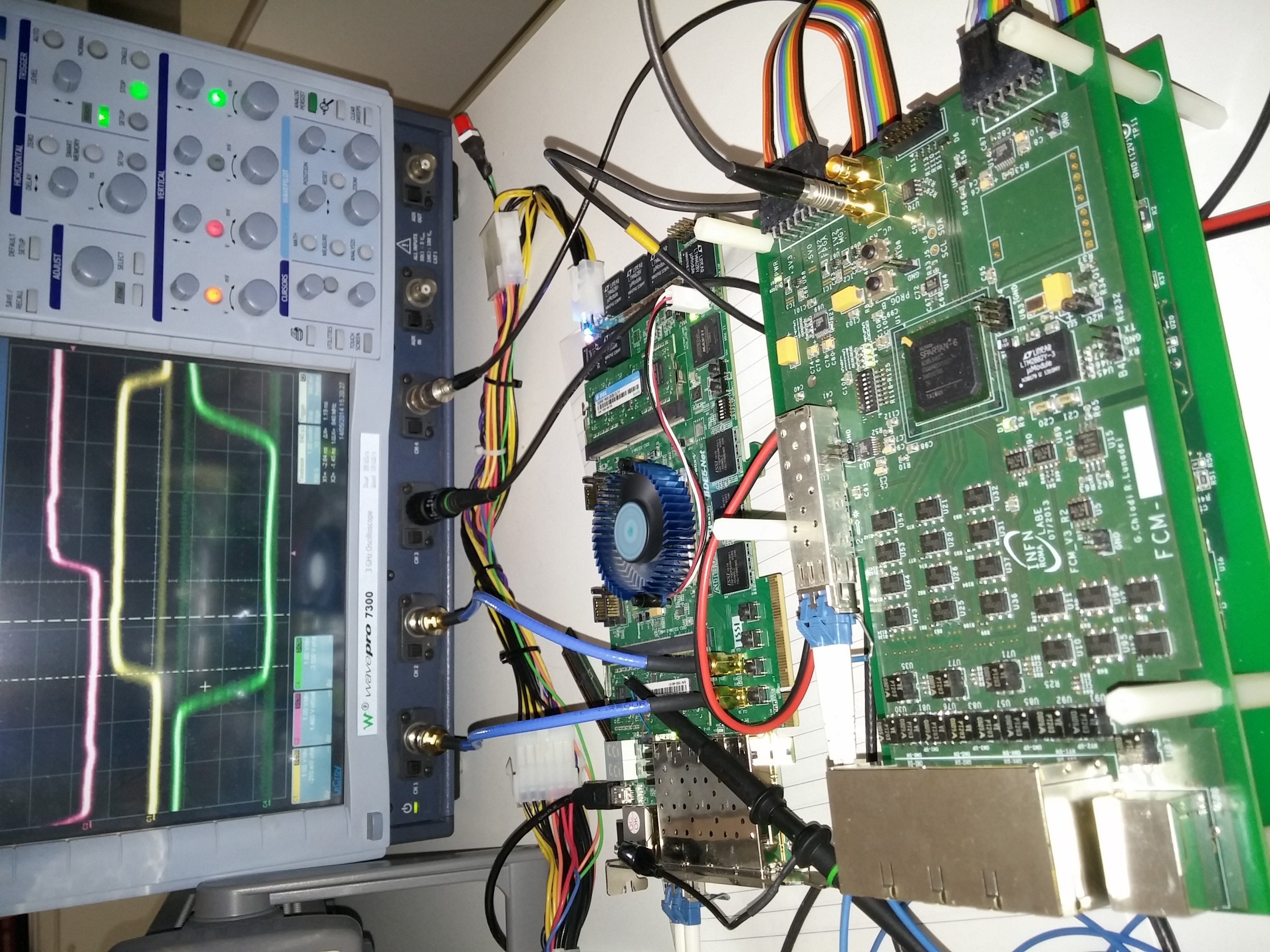}
  \caption{\nanetcube testbed.}
  \label{fig:testbed}
\end{figure}

Preliminary results show that the interoperability between different
vendors FPGA devices can be achieved and the timing resolution
complies with physics requirements.\\
We develop a test setup to explore the fixed latency capabilities of a
complete links chain.\\
We leverage on the fixed latency native mode of the \altera
transceivers and on the hardware fixed latency implementation for
Xilinx device~\cite{Giordano:2011}.
The testbed is composed by the \nanetcube board and the FCM
\mbox{Xilinx-based} board respectively emulating the \mbox{on-shore}
and \mbox{off-shore} boards connected by optical fibers (see
Fig.~\ref{fig:testbed}).\\
The external \mbox{GPS-equivalent} clock has been input to the
\nanetcube to clock the transmitting side of the device.
A sequence of dummy parallel data are serialised, 8b/10b encoded and
transmitted, together with the embedded serial clock, at a data rate
of 800~Mbps along the fiber towards the receiver side of the FCM
system.
The FCM system recovers from the received clock and transmits the
received data and recovered clock back to the \nanetcube boards.
Lastly, the received side of \nanetcube deserializes data and produces
the received clock.\\

\begin{figure}[!htb]
  \centering
  \includegraphics[width=.49\textwidth]{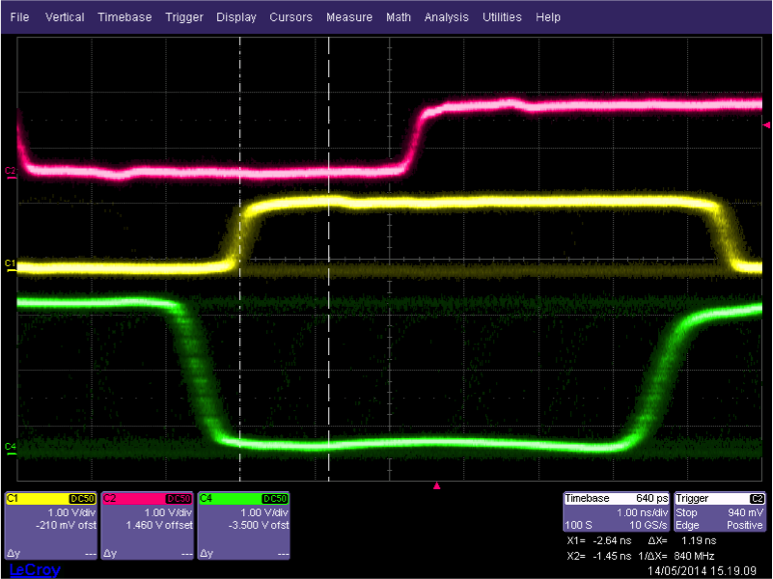}
  \caption{Deterministic latency feature of \nanetcube SerDes: the
    plot scope shows the phase alignment of the transmitting (purple)
    and receiving (yellow) parallel clocks after 12~h test of periodic
    reset and initialisation sequence.}
  \label{fig:Fixed_Lat}
\end{figure}

The way to test the fixed latency features of the SerDes hardware
implementation is quite easy taking into account that every time a new
initialisation sequence, following an hardware reset or a powerup of
the SerDes hardware, has been done, we should be able to measure the
same phase shift between transmitted and received clock, equal to the
fixed number of serial clock cycles shift used to correctly align the
deserialised data stream.
Fig.~\ref{fig:Fixed_Lat} is a picture taken from scope acquisition in
Infinity Persistence showing the results of a preliminary 12~h test
where every 10~s a new \emph{reset and align} procedure has been
issued.
The \nanetcube transmitter parallel clock (the purple signal)
maintains exactly the same phase difference with the receiver parallel
clock (the yellow signal) and with the FCM recovered clock (the green
signal).

\section{Conclusions and future work} 
Our \nanet design proved to be efficient in performing \realtime data
communication between the NA62 RICH readout system and the GPU-based L0
trigger processor over a single GbE link.
Preliminary results of its customization for the data transport system
of the \km experiment shows that the fundamental requirement of a
deterministic latency link can be implemented using \nanet, paving the
way to the use of hybrid trigger and data reconstruction systems.
\nanetten 10 GbE board, currently under development, will allow for a
full integration of our architecture in the NA62 experiment and smooth
the path to \nanet usage in other contexts.




%




\end{document}